\documentclass[12pt]{article}

\def\proof{\medbreak\noindent{\bf Proof}}

\def\theorem #1. #2\par{\medbreak
  \noindent{\tt {\bf Theorem #1.}\enspace}{\sl#2\par}%
  \ifdim\lastskip<\medskipamount \removelastskip\penalty55\medskip\fi}

\def\for{\quad\hbox{for}\quad}
\def\n{{\scriptsize\textsc n}}
\def\k{{\scriptsize\textsc k}}

\def\ww{\wedge\ldots\wedge}
\def\la{\Lambda}

\def\I{{\cal I}}
\def\R{{\cal R}}
\def\M{{\cal M}}
\def\C{{\cal C}}
\def\L{{\cal L}}
\def\W{{\cal W}}
\def\D{{\cal D}}
\def\K{{\cal K}}
\def\Tr{{\rm Tr}}
\def\Spin{{\rm Spin}}
\def\diag{{\rm diag}}
\def\even{{\rm even}}
\def\odd{{\rm odd}}
\def\det{{\rm det}}
\def\SO{{\rm SO}}
\def\exp{{\rm exp}}

\def\be{\begin{equation}}
\def\ee{\end{equation}}

\begin{document}

\author{N.G.Marchuk \thanks{Research supported by the Russian Foundation
for Basic Research grants 00-01-00224,  00-15-96073.}}

\title{A concept of Dirac-type tensor equations}

\maketitle


\vskip 1cm

Steklov Mathematical Institute, Gubkina st.8, Moscow 119991, Russia

nmarchuk@mi.ras.ru,   www.orc.ru/\~{}nmarchuk
\vskip 1cm

\begin{abstract}
Considering a four dimensional parallelisable manifold,
we develop a concept
of Dirac-type tensor equations with wave functions that belong to left
ideals of the set of nonhomogeneous complex valued differential forms.
\end{abstract}

\tableofcontents
\bigskip


In the previous papers \cite{nona},\cite{nona1},\cite{nona2},
developing ideas of
\cite{Ivanenko}-\cite{Benn}, we consider Dirac-type
tensor equations with non-Abelian gauge symmetries on a four
dimensional parallelisable
manifold with wave functions that belong to the algebrae
$\Lambda^\C,\Lambda,\Lambda^\C_\even,\Lambda_\even$
of real dimensions $32,16,16,8$ respectively,
where $\Lambda$ is the set of nonhomogeneous differential forms,
$\Lambda_\even$ is the set of even differential forms, and
$\Lambda^\C,\Lambda^\C_\even$ are the corresponding sets of complex
valued differential forms. Now we develop a concept
of Dirac-type tensor equations with wave functions that belong to left
ideals of $\Lambda^\C$. This concept gives us possibility to find an
${\rm SU}(3)$ invariant
Dirac-type tensor equation with 12 complex valued components
of wave function.
In addition to unitary gauge symmetries Dirac-type tensor equations have a gauge
symmetry with respect to the spinor group $\Spin(\W)$. In Section 9 we
investigate in detail a connection between the (spinor) Dirac equation
and the Dirac-type tensor equation. Of cause, Dirac spinors and tensors  are
different mathematical objects and, generally speaking, we can't establish
an equivalence between them. But it nevertheless, if we take Dirac spinors
that are solutions of the Dirac equation and tensors (differential forms) that
are solutions of the corresponding Dirac-type tensor equation, then, with
the aid of above mentioned spinor group symmetry, we can establish a
one-to-one correspondence between these spinors and tensors (Section 9).


\section{Two pictures of the Dirac equation.}
Let $\R^{1,3}$ be the Minkowski space with coordinates $x^\mu$ ($\mu=0,1,2,3$), with
a metric tensor $\eta_{\mu\nu}=\eta^{\mu\nu}$ defined by the Minkowski matrix
$$
\eta=\|\eta_{\mu\nu}\|=\|\eta^{\mu\nu}\|=\diag(1,-1,-1,-1).
$$

Consider the Dirac equation for an electron
\begin{equation}
\gamma^\mu(\partial_\mu\psi+ia_\mu\psi)+im\psi=0,
\label{Dirac:eq}
\end{equation}
where $\partial_\mu=\partial/\partial x^\mu$,
$\psi=\psi(x)$ is a column of four complex valued
smooth functions (an electron wave function), $a_\mu=a_\mu(x)$ is a real valued covector
field (a potential of electromagnetic field), $m$ is a real constant (the electron mass),
$i=\sqrt{-1}$, and $\gamma^\mu$ are (Dirac) matrices that satisfy the identities
(${\bf 1}$ is the identity matrix)
$$
\gamma^\mu\gamma^\nu+\gamma^\nu\gamma^\mu=2\eta^{\mu\nu}{\bf 1}.
$$
In particular we may take
\begin{eqnarray}
\gamma^0&=&\pmatrix{1 &0 &0 & 0\cr
                  0 &1 & 0&0 \cr
                  0 &0 &-1&0 \cr
                  0 &0 &0 &-1},\quad
\gamma^1=\pmatrix{0 &0 &0 & 1\cr
                  0 &0 & 1&0 \cr
                  0 &-1 &0 &0 \cr
                  -1 &0 &0 &0},
\label{gamma:matrices}\\
\gamma^2&=&\pmatrix{0 &0 &0 & -i\cr
                  0 &0 & i&0 \cr
                  0 & i&0 &0 \cr
                  -i &0 &0 &0},\quad
\gamma^3=\pmatrix{0 &0 & 1& 0\cr
                  0 &0 & 0&-1 \cr
                  -1 &0 &0 &0 \cr
                  0 &1&0 &0}.\nonumber
\end{eqnarray}
Consider a change of coordinates
\begin{equation}
x^\mu\to\acute x^\mu=p^\mu_\nu x^\nu
\label{change:coord}
\end{equation}
from the proper orthochroneous Lorentz group $\SO^+(1,3)$, i.e.,
\begin{equation}
P^{\rm T}\eta P=\eta,\quad \det P=1,\quad p^0_0>0,
\label{proper}
\end{equation}
where $P=\|p^\mu_\nu\|$. Then
$$
\partial_\mu\to\acute\partial_\mu=q^\nu_\mu\partial_\nu,\quad
a_\mu\to\acute a_\mu=q^\nu_\mu a_\nu,
$$
where $q^\nu_\mu$ are elements of the inverse matrix $P^{-1}$, i.e.,
$$
q^\nu_\mu p^\mu_\lambda=\delta^\nu_\lambda,\quad
q^\mu_\nu p^\lambda_\mu=\delta^\lambda_\nu
$$
and $\delta^\mu_\nu$ is the Kronecker tensor.
P.~Dirac in \cite{Dirac} postulate that $\gamma^\mu$ is a vector with
values in $\M^\C(4)$ ($\M^\C(4)$ is the algebra of complex valued $4\!\times\!4$-matrices).
That means
$$
\gamma^\mu\to\acute\gamma^\mu=p^\mu_\nu\gamma^\nu
$$
and eq. (\ref{Dirac:eq}) in coordinates $(\acute x)$ has the form
\begin{equation}
\acute\gamma^\mu(\acute\partial_\mu\psi+i\acute a_\mu\psi)+im\psi=0,
\label{Dirac:eq:prime}
\end{equation}
where $\acute\partial_\mu=\partial/\partial\acute x^\mu$,
$\psi=\psi(x(\acute x))$. It is proved in the theory of representations of
Lorentz group that there exists a pair of nondegenerate matrices $\pm R\in\M^\C(4)$
such that
\begin{equation}
R^{-1}\gamma^\mu R=p^\mu_\nu\gamma^\nu=\acute \gamma^\mu.
\label{gamma:prime}
\end{equation}
Substituting $R^{-1}\gamma^\mu R$ for $\acute\gamma^\mu$ in (\ref{Dirac:eq:prime}),
we get
\begin{equation}
R^{-1}\gamma^\mu R(\acute\partial_\mu\psi+i\acute a_\mu\psi)+im\psi=0,
\label{Dirac:eq:prime2}
\end{equation}
or, equivalently,
\begin{equation}
\gamma^\mu(\acute\partial_\mu(R\psi)+i\acute a_\mu(R\psi))+im(R\psi)=0,
\label{Dirac:eq:spinor}
\end{equation}
We see that an invariance of the Dirac equation can be proved in two ways.

If we suppose the following transformation rules for $\gamma^\mu$ and $\psi$ under
every change of coordinates (\ref{change:coord},\ref{proper}):
$$
\gamma^\mu\to R^{-1}\gamma^\mu R,\quad \psi\to\psi,
$$
then we get {\em a spinorless picture} of the Dirac equation.

In the case of transformation rules
$$
\gamma^\mu\to\gamma^\mu,\quad \psi\to R\psi,
$$
we get a conventional  {\em spinor picture} of the Dirac equation.
In sect.9, considering a connection between the Dirac equation and the
corresponding Dirac-type tensor equation, we shall see that the two
pictures of the Dirac equation are the consequence of a gauge symmetry
of the Dirac-type tensor equation with respect to the spinor group.

\section{A Differentiable manifold $X$. Tensors.}

Let $X$ be a four-dimensional orientable
differentiable manifold with local coordinates
$x^\mu$ ($\mu=0,1,2,3$). Summation convention over repeating indices is assumed.
Let $\top^q_p$ be the set of smooth type $(p,q)$ real valued tensors (tensor fields) on $X$
$$
\top^q_p=\{u^{\nu_1\ldots\nu_q}_{\mu_1\ldots\mu_p}=
u^{\nu_1\ldots\nu_q}_{\mu_1\ldots\mu_p}(x)\}.
$$
Under a change of coordinates $(x)\to(\acute x)$ a tensor
$u^{\nu_1\ldots\nu_q}_{\mu_1\ldots\mu_p}$ transforms as follows:
$$
u^{\nu_1\ldots\nu_q}_{\mu_1\ldots\mu_p}\to
\acute u^{\beta_1\ldots\beta_q}_{\alpha_1\ldots\alpha_p}=
u^{\nu_1\ldots\nu_q}_{\mu_1\ldots\mu_p}
\frac{\partial\acute x^{\beta_1}}{\partial x^{\nu_1}}\ldots
\frac{\partial\acute x^{\beta_q}}{\partial x^{\nu_q}}
\frac{\partial x^{\mu_1}}{\partial\acute x^{\alpha_1}}\ldots
\frac{\partial x^{\mu_p}}{\partial\acute x^{\alpha_p}}.
$$
Operations of addition and tensor product are defined in the usual way
$$
(u+v)^{\nu_1\ldots\nu_q}_{\mu_1\ldots\mu_p}=
u^{\nu_1\ldots\nu_q}_{\mu_1\ldots\mu_p}+v^{\nu_1\ldots\nu_q}_{\mu_1\ldots\mu_p}
\in\top^q_p,\quad u,v\in\top^q_p;
$$
$$
(u\otimes v)^{\nu_1\ldots\nu_{q+s}}_{\mu_1\ldots\mu_{p+r}}=
u^{\nu_1\ldots\nu_q}_{\mu_1\ldots\mu_p}
v^{\nu_{q+1}\ldots\nu_{q+s}}_{\mu_{p+1}\ldots\mu_{p+r}}\in\top^{q+s}_{p+r},\quad
u\in\top^q_p,v\in\top^s_r.
$$


\section{A Parallelisable manifold with a tetrad.}
An $n$-dimensional differentiable manifold is called {\em parallelisable} if there
exist $n$ linear independent vector or covector fields on it. Let $X$ be a four
dimensional parallelisable manifold with local coordinates $x=(x^\mu)$ and
$$
e_\mu{}^a=e_\mu{}^a(x),\quad a=0,1,2,3
$$
be four covector fields on $X$. This set of four covectors is called {\em a tetrad}.
The full set $\{X,e_\mu{}^a,\eta\}$, where $\eta$ is the Minkowski matrix, is denoted
by $\W$. Here and in what follows we use Greek indices as tensorial indices and Latin
indices as nontensorial (tetrad) indices.

We can define a metric tensor on $W$
$$
g_{\mu\nu}=e_\mu{}^a e_\nu{}^b\eta_{ab}
$$
such that
$$
g_{\mu\nu}=g_{\nu\mu},\quad g_{00}>0,\quad  g=\det\|g_{\mu\nu}\|<0,
$$
and the signature of the matrix $\|g_{\mu\nu}\|$ is equal to $-2$. Hence we may
consider $\W$ as pseudo-Riemannian space with the metric tensor $g_{\mu\nu}$.
We raise and lower Latin indices with the aid of the Minkowski matrix
$\eta^{ab}=\eta_{ab}$
and Greek indices with the aid of the metric tensor $g_{\mu\nu}$
$$
e^{\nu a}=g^{\mu\nu}e_\mu{}^a,\quad e_{\mu a}=\eta_{ab}e_\mu{}^b.
$$
With the aid of the tetrad we can replace all or part of Greek indices in a tensor
$u^{\nu_1\ldots\nu_q}_{\mu_1\ldots\mu_p}\in\top^q_p$ by Latin indices
$$
u^{b_1\ldots b_q}_{a_1\ldots a_p}=
u^{\nu_1\ldots\nu_q}_{\mu_1\ldots\mu_p} e^{\mu_1}{}_{a_1}\ldots
e^{\mu_p}{}_{a_p}e_{\nu_1}{}^{b_1}\ldots e_{\nu_q}{}^{b_q}.
$$
As a result we get a set of invariants $u^{b_1\ldots b_q}_{a_1\ldots a_p}$
enumerated by Latin indices. On the contrary we can replace Latin indices by
Greek indices
$$
u^{\nu_1\ldots\nu_q}_{\mu_1\ldots\mu_p}=
u^{b_1\ldots b_q}_{a_1\ldots a_p} e^{\nu_1}{}_{b_1}\ldots
e^{\nu_p}{}_{b_p}e_{\mu_1}{}^{a_1}\ldots e_{\mu_q}{}^{a_q}.
$$
Note that
$$
e^\mu{}_a e_\mu{}^b=\delta^b_a,\quad
e^\mu{}_a e_\nu{}^a=\delta^\mu_\nu.
$$
The metric tensor $g_{\mu\nu}$ defines the Levi-Civita connection, the curvature
tensor, the Ricci tensor, and the scalar curvature
\begin{eqnarray}
{\Gamma_{\mu\nu}}^\lambda&=&
\frac{1}{2}g^{\lambda\kappa}(\partial_\mu g_{\nu\kappa}+
\partial_{\nu}g_{\mu\kappa}-\partial_\kappa g_{\mu\nu}),
\label{Levi-Civita}\\
{R}_{\lambda\mu\nu}{}^\kappa&=&
\partial_\mu {\Gamma}_{\nu\lambda}{}^\kappa-\partial_\nu
{\Gamma}_{\mu\lambda}{}^\kappa+
{\Gamma}_{\mu\eta}{}^\kappa{\Gamma}_{\nu\lambda}{}^\eta-
{\Gamma}_{\nu\eta}{}^\kappa{\Gamma}_{\mu\lambda}{}^\eta,
\label{curv}\\
R_{\nu\rho}&=&{R^\mu}_{\nu\mu\rho},
\label{Ricci}\\
R&=&g^{\rho\nu}R_{\rho\nu}
\label{scalar-curv}
\end{eqnarray}
with symmetries
$$
{\Gamma_{\mu\nu}}^\lambda={\Gamma_{\nu\mu}}^\lambda,\quad
R_{\mu\nu\lambda\rho}=R_{\lambda\rho\mu\nu}=R_{[\mu\nu]\lambda\rho},\quad
R_{\mu[\nu\lambda\rho]}=0,\quad
R_{\nu\rho}=R_{\rho\nu}
$$
Covariant derivatives $\nabla_\mu\,:\,\top^q_p\to\top^q_{p+1}$ are defined
via the Levi-Civita connection by the usual rules
\medskip

\noindent 1. If $u=u(x),\ x\in X$ is a scalar function, then
$$
\nabla_\mu u=\partial_\mu u.
$$

\noindent 2. If $u^\nu\in\top^1$, then
$$
\nabla_\mu u^\nu\equiv u^\nu_{;\mu}=\partial_\mu u^\nu +
{\Gamma_{\lambda\mu}}^\nu u^\lambda.
$$

\noindent 3. If $u_\nu\in\top_1$, then
$$
\nabla_\mu u_\nu\equiv u_{\nu;\mu}=\partial_\mu u_\nu -
{\Gamma_{\nu\mu}}^\lambda u_\lambda.
$$

\noindent 4. If $u=(u^{\nu_1\ldots \nu_k}_{\lambda_1\ldots \lambda_l})\in\top^k_l$,
$v=(v^{\mu_1\ldots \mu_r}_{\rho_1\ldots \rho_s})\in\top^r_s$, then
$$
\nabla_\mu(u\otimes v)=(\nabla_\mu u)\otimes v + u\otimes\nabla_\mu v.
$$
\medskip

With the aid of these rules it is easy to calculate covariant derivatives of
arbitrary type tensors. Also we can check the formulas
\begin{eqnarray*}
&&\nabla_\mu g_{\alpha\beta}=0,\quad
\nabla_\mu g^{\alpha\beta}=0,\quad
\nabla_\mu \delta_\alpha^\beta=0,\\
&&\nabla_\alpha(R^{\alpha\beta}-\frac{1}{2}R g^{\alpha\beta})=0,\\
&&(\nabla_\mu \nabla_\nu-\nabla_\nu \nabla_\mu)a_\rho=
{R^\lambda}_{\rho\mu\nu}a_\lambda,
\end{eqnarray*}
for any $a_\rho\in\top_1$.


\section{Differential forms.}
Let $\Lambda_k$ be the set of all exterior differential forms of rank $k=0,1,2,3,4$
($k$-forms) on $\W$ and
$$
\Lambda=\Lambda^0\oplus\ldots\oplus\Lambda^4=\Lambda^\even\oplus\Lambda^\odd,
$$
$$
\Lambda^\even=\Lambda^0\oplus\Lambda^2\oplus\Lambda^4,\quad
\Lambda^\odd=\Lambda^1\oplus\Lambda^3.
$$
The set of smooth real valued functions on $\W$ is identified with the set of
0-forms $\Lambda_0$. A $k$-form $U\in\Lambda_k$ can be written as
\begin{equation}
U=\frac{1}{k!} u_{\nu_1\ldots \nu_k}dx^{\nu_1}\ww dx^{\nu_k}=
\sum_{\mu_1<\cdots<\mu_k} u_{\mu_1\ldots \mu_k}dx^{\mu_1}\ww dx^{\mu_k},
\label{k-form}
\end{equation}
where $u_{\nu_1\ldots \nu_k}=u_{\nu_1\ldots \nu_k}(x)$ are real valued
components of a covariant antisymmetric
($u_{\nu_1\ldots \nu_k}=u_{[\nu_1\ldots \nu_k]}$) tensor.
Differential forms from $\Lambda$ can be written as linear combinations of
the 16 basis elements
\begin{equation}
1,dx^\mu,dx^{\mu_1}\wedge dx^{\mu_2},\ldots,dx^{0}\wedge\ldots\wedge
dx^3,
\quad
\mu_1<\mu_2<\ldots.
\label{basis}
\end{equation}
The exterior product of differential forms is defined in the
usual way.
If $U\in\Lambda^r,V\in\Lambda^s$, then
$$
U\wedge V=(-1)^{rs}V\wedge U\in\la^{r+s}.
$$
In the sequel we also use complex valued differential forms from
$\Lambda^\C_k,\Lambda^\C,\Lambda^\C_\even,\Lambda^\C_\odd$.

The tetrad $e_\mu{}^a$ can be written with the aid of four 1-forms
$$
e^a=e_\mu{}^a dx^\mu\quad \hbox{or}\quad e_a=\eta_{ab}e^b.
$$
The $k$-form $U$ from (\ref{k-form}) can be written as
$$
U=\frac{1}{k!}u_{a_1\ldots a_k}e^{a_1}\ww e^{a_k},
$$
where invariants $u_{a_1\ldots a_k}=u_{[a_1\ldots a_k]}$
connected with tensor components $u_{\mu_1\ldots \mu_k}$ by the formula
$$
u_{a_1\ldots a_k}=u_{\mu_1\ldots \mu_k}e^{\mu_1}{}_{a_1}\ldots e^{\mu_k}{}_{a_k}.
$$
Differential forms from $\Lambda$ can be written as
$$
\sum_{k=0}^4 \frac{1}{k!}u_{a_1\ldots a_k}e^{a_1}\ww e^{a_k},\quad
u_{a_1\ldots a_k}=u_{[a_1\ldots a_k]}
$$
and the 16 differential forms
\begin{equation}
1,e^a,e^{a_1}\wedge e^{a_2},\ldots e^0\ww e^3,\quad
a_1<a_2<\ldots
\label{tetrad:basis}
\end{equation}
can be considered as basis forms of $\Lambda$.


\section{A central product of differential forms.}
Let us define {\em a central product} of differential forms
$U,V\to UV$ by the following rules:

\noindent 1. For $U,V,W\in\Lambda,\,\alpha\in\Lambda_0$
$$
1U=U1=U,\quad
(\alpha U)V=U(\alpha V)=\alpha(UV),
$$
$$
U(VW)=(UV)W,\quad
(U+V)W=UW+VW.
$$

\noindent 2. $e^a e^b=e^a\wedge e^b+\eta^{ab}$.

\noindent 3. $e^{a_1}\ldots e^{a_k}=e^{a_1}\ww e^{a_k}$ for $a_1<\ldots<a_k$.

\medskip
Note that from the second rule we get the equality
\begin{equation}
e^a e^b+e^b e^a=2\eta^{ab},
\label{Clifford:rel}
\end{equation}
which appear in the Clifford algebra\footnote{In a special case the central product
was invented by H.~Grassmann \cite{Grassmann} in 1877 as an attempt to unify the
exterior calculus (the Grassmann algebra) with the quaternion calculus. A discussion
on that matter see in \cite{Doran}. In some papers the central product is called
a Clifford product.}. Substituting into (\ref{Clifford:rel})
$$
e^a=e_\mu{}^a dx^\mu,\quad
e^b=e_\nu{}^b dx^\nu,\quad
\eta^{ab}=g^{\mu\nu}e_\mu{}^a e_\nu{}^b,
$$
we get
$$
e_\mu{}^a e_\nu{}^b(dx^\mu dx^\nu+dx^\nu dx^\mu-2g^{\mu\nu})=0.
$$
Thus,
$$
dx^\mu dx^\nu+dx^\nu dx^\mu=2g^{\mu\nu}.
$$
Evidently, the operation of central product maps $\Lambda\times\Lambda\to\Lambda$.
From this definition of the central product it is evident that for
$u_{a_1\ldots a_k}=u_{[a_1\ldots a_k]}$
$$
\frac{1}{k!}u_{a_1\ldots a_k}e^{a_1}\ww e^{a_k}=
\frac{1}{k!}u_{a_1\ldots a_k}e^{a_1}\ldots e^{a_k}.
$$

In the sequel we use notation
$$
e^{a_1\ldots a_k}=e^{a_1}\ldots e^{a_k}=e^{a_1}\ww e^{a_k}\for
a_1<\cdots<a_k.
$$

Let us define an operation of conjugation, which maps
$\Lambda_k\to\Lambda_k$ or $\Lambda_k^\C\to\Lambda_k^\C$. For
$U\in\Lambda_k^\C$
$$
U^*=(-1)^{\frac{k(k-1)}{2}}\bar{U},
$$
where $\bar{U}$ is the differential form with complex conjugated
components (if $U\in\Lambda$, then $\bar{U}=U$). We see that
$$
(UV)^*=V^*U^*,\quad U^{**}=U,\quad
(e^{a_1}\ldots e^{a_k})^*=e^{a_k}\ldots e^{a_1}
$$
and
$$
U^*=U\quad\hbox{for}\quad U\in\Lambda_0\oplus\Lambda_1\oplus\Lambda_4,
$$
$$
U^*=-U\quad\hbox{for}\quad U\in\Lambda_2\oplus\Lambda_3.
$$

Let us define {\em a trace} of differential form as linear operation
$\Tr:\Lambda\to\Lambda_0$ such that
$$
\Tr(1)=1,\quad \Tr(e^{a_1}\ww e^{a_k})=0\quad\hbox{for}\quad
k=1,2,3,4.
$$
It is easy to prove that
$$
\Tr(UV-VU)=0\quad\hbox{for}\quad U,V\in\Lambda.
$$

Let us take the set of forms
$$
\Spin(\W)=\{S\in\Lambda_\even\,:\,S^*S=1\},
$$
which can be considered as a Lie group with respect to the central
product. It can be shown that the real Lie algebra of the Lie group
$\Spin(\W)$ is coincide with the set of 2-forms $\Lambda_2$ with
the commutator $[U,V]=UV-VU$, which maps
$\Lambda_2\times\Lambda_2\to\Lambda_2$. If $U\in\Lambda_2$, then
$\pm\exp(U)\in\Spin(\W)$, where
$$
\exp(U)=1+\sum_{k=1}^\infty\frac{1}{k!}U^k.
$$
In some cases it is suitable to use for calculations {\em an exterior
exponent} \cite{Lounesto}. If $u_{a_1 a_2}=u_{[a_1 a_2]}$ and
$U=\frac{1}{2} u_{a_1 a_2}e^{a_1}\wedge e^{a_2}\in\Lambda_2$ is such that
\begin{eqnarray*}
\lambda&=&1 - u_{01}{}^2 - u_{02}{}^2 - u_{03}{}^2 + u_{12}{}^2 -
      u_{03}{}^2 u_{12}{}^2 + 2 u_{02} u_{03} u_{12} u_{13} \\
    && + u_{13}{}^2 - u_{02}{}^2 u_{13}{}^2 -
      2 u_{01} u_{03} u_{12} u_{23} +
      2 u_{01} u_{02} u_{13} u_{23} + u_{23}{}^2 -
      u_{01}{}^2 u_{23}{}^2>0,
\end{eqnarray*}
then
$$
\pm\frac{1}{\sqrt{\lambda}}(1+U+\frac{1}{2}U\wedge U)\in\Spin(\W).
$$
It is not hard to prove that if $U\in\Lambda_k$ and $S\in\Spin(\W)$,
then
$$
S^{-1}US\in\Lambda_k,
$$
i.e., $S^{-1}\Lambda_k S\subseteq\Lambda_k$.
In particular,
$$
S^{-1}e^a S=p^a_b e^b,
$$
where the matrix $P=\|p^a_b\|$ has the properties
$$
P^{\rm T}\eta P=\eta,\quad\det P=1,\quad p^0_0>0.
$$
The transformation
\begin{equation}
e^a\to \check e^a=S^{-1}e^a S,
\label{rot:tetrad}
\end{equation}
where $S\in\Spin(\W)$, is called {\em a Lorentz rotation of the tetrad}.
Obviously,
\begin{eqnarray*}
S^{-1}(\frac{1}{k!}u_{a_1\ldots a_k}e^{a_1}\ww e^{a_k})S&=&
S^{-1}(\frac{1}{k!}u_{a_1\ldots a_k}e^{a_1}\ldots e^{a_k})S\\
&=&\frac{1}{k!}u_{a_1\ldots a_k}(S^{-1}e^{a_1}S)\ldots(S^{-1} e^{a_k}S)\\
&=&\frac{1}{k!}u_{a_1\ldots a_k}\check e^{a_1}\ldots\check e^{a_k}
=\frac{1}{k!}u_{a_1\ldots a_k}\check e^{a_1}\ww\check e^{a_k}.
\end{eqnarray*}
From the identities $e^a e^b+e^b e^a=2\eta^{ab}$ it follows that
$$
\check e^a\check e^b+\check e^b\check e^a=2\eta^{ab}.
$$

\section{Tensors from $\Lambda_k\top_s^r$.}
Let us take a tensor
$$
u^{\lambda_1\ldots\lambda_r}_{\nu_1\ldots\nu_s\mu_1\ldots\mu_k}=
u^{\lambda_1\ldots\lambda_r}_{\nu_1\ldots\nu_s[\mu_1\ldots\mu_k]}
\in\top^r_{s+k}
$$
antisymmetric with respect to $k$ covariant indices. One may consider
the following objects:
\begin{equation}
U^{\lambda_1\ldots\lambda_r}_{\nu_1\ldots\nu_s}=
\frac{1}{k!}u^{\lambda_1\ldots\lambda_r}_{\nu_1\ldots\nu_s\mu_1\ldots\mu_k}
dx^{\mu_1}\ww dx^{\mu_k},
\label{lambda:tensor}
\end{equation}
which are formally written as $k$-forms. Under a change of coordinates
$(x)\to(\acute x)$ values (\ref{lambda:tensor}) transform as components
of tensor of type $(s,r)$, i.e.,
$$
\acute U^{\alpha_1\ldots\alpha_r}_{\beta_1\ldots\beta_s}=
q^{\nu_1}_{\beta_1}\ldots q^{\nu_s}_{\beta_s}
p^{\alpha_1}_{\lambda_1}\ldots
p^{\alpha_r}_{\lambda_r}
U^{\lambda_1\ldots\lambda_r}_{\nu_1\ldots\nu_s},
\quad q^\nu_\beta=\frac{\partial x^\nu}{\partial\acute x^\beta},\quad
p^\alpha_\lambda=\frac{\partial\acute x^\alpha}{\partial x^\lambda}.
$$
The objects (\ref{lambda:tensor}) are called
{\em tensors of type $(s,r)$ with values in $\Lambda_k$}. We write this
as
$$
U^{\lambda_1\ldots\lambda_r}_{\nu_1\ldots\nu_s}\in\Lambda_k\top^r_s.
$$
We take
$$
\Lambda\top^r_s=\Lambda_0\top^r_s\oplus\ldots\oplus \Lambda_4\top^r_s.
$$

Note that $dx^\mu=\delta^\mu_\nu dx^\nu$, where $\delta^\mu_\nu$ is the
Kronecker tensor ($\delta^\mu_\nu=0$ for $\mu\neq\nu$ and
$\delta^\mu_\mu=1$). Hence, $dx^\mu\in\Lambda_1\top^1$.

Elements of $\Lambda_0\top^r_s$ are identified with tensors from
$\top^r_s$. For the sequel it is suitable to write tensors
(\ref{lambda:tensor}) as
$$
U^{\lambda_1\ldots\lambda_r}_{\nu_1\ldots\nu_s}=
\frac{1}{k!}
u^{\lambda_1\ldots\lambda_r}_{\nu_1\ldots\nu_s a_1\ldots a_k}
e^{a_1}\ww e^{a_k}\in\Lambda_k\top^r_s.
$$
Let us define a central product of elements
$U^{\mu_1\ldots\mu_r}_{\nu_1\ldots\nu_s}\in\Lambda\top^r_s$
and
$V^{\alpha_1\ldots\alpha_p}_{\beta_1\ldots\beta_q}\in\Lambda\top^p_q$ as
a tensor from $\Lambda\top^{r+p}_{s+q}$ of the form
$$
W^{\mu_1\ldots\mu_r\alpha_1\ldots\alpha_p}_{\nu_1\ldots\nu_s\beta_1\ldots\beta_q}
=U^{\mu_1\ldots\mu_r}_{\nu_1\ldots\nu_s}
V^{\alpha_1\ldots\alpha_p}_{\beta_1\ldots\beta_q},
$$
where on the right-hand side there is the central product of
differential forms (the indices
$\mu_1,\ldots,\mu_r,\alpha_1,\ldots,\alpha_p,\nu_1,\ldots,\nu_s,\beta_1,\ldots,\beta_q$
are fixed).

If $U^{\mu_1\ldots\mu_r}_{\nu_1\ldots\nu_s}\in\Lambda_0\top^r_s$
and
$V^{\alpha_1\ldots\alpha_p}_{\beta_1\ldots\beta_q}\in\Lambda_0\top^p_q$,
then the central product of these elements is identified with the tensor
product.


Let us define operators (Upsilon) $\Upsilon_\mu$, which act on tensors
from $\Lambda\top^r_s$ by the following rules:
\medskip

\noindent a) If $u=(u_{\nu_1\ldots \nu_s}^{\epsilon_1\ldots \epsilon_r})
\in\Lambda_0\top^r_s$, then
$$
\Upsilon_\mu u=\partial_\mu u.
$$

\noindent b) $\Upsilon_\mu dx^\nu = -{\Gamma^\nu}_{\mu\lambda}
dx^\lambda$.
\smallskip

\noindent c)  If $U\in\Lambda\top^r_s$, $V\in\Lambda\top^p_q$, then
$$
\Upsilon_\lambda(UV)=
(\Upsilon_\lambda U)V+U\Upsilon_\lambda V.
$$
\smallskip

\noindent d)  If $U,V\in\Lambda\top^r_s$, then
$$
\Upsilon_\lambda(U+V)=\Upsilon_\lambda U+\Upsilon_\lambda V.
$$
\medskip

With the aid of these rules it is easy to calculate how operators
$\Upsilon_\mu$ act on arbitrary tensor from
$\Lambda\top_s^r$.

\theorem 1. If $U\in\Lambda_k$ has the form (\ref{k-form}), then
$$
\Upsilon_\nu U=\frac{1}{k!}u_{\mu_1\ldots\mu_k;\nu}dx^{\mu_1}\ww dx^{\mu_k}\in
\Lambda_k\top_1.
$$
\par

\noindent The proof in straightforward.
\medskip

From the rule b) we get
\begin{equation}
(\Upsilon_\mu\Upsilon_\nu-\Upsilon_\nu\Upsilon_\mu)dx^\lambda=
-{R^\lambda}_{\rho\mu\nu}dx^\rho.
\label{R1}
\end{equation}

\theorem 2. Under a change of coordinates $(x)\to(\acute x)$ operators $\Upsilon_\mu$
transform as components of a covector, i.e.,
$$
\acute\Upsilon_\mu=q^\nu_\mu \Upsilon_\nu,\quad
q^\nu_\mu=\frac{\partial x^\nu}{\partial\acute x^\mu}.
$$

\proof. This fact follows from the transformation rule of Christoffel symbols
$\Gamma_{\mu\nu}{}^\lambda$.
\medskip


\theorem 3.
$$
e^a U e_a=\left\{\begin{array}{ll}
4U&\mbox{for $U\in\Lambda_0\top^p_q$}\\
-2U&\mbox{for $U\in\Lambda_1\top^p_q$}\\
0&\mbox{for $U\in\Lambda_2\top^p_q$}\\
2U&\mbox{for $U\in\Lambda_3\top^p_q$}\\
-4U&\mbox{for $U\in\Lambda_4\top^p_q$}
\end{array}
\right.
$$

The proof is by direct calculations.
\medskip

Let us take the tensor
\begin{equation}
B_\mu=-\frac{1}{4}e^a\wedge\Upsilon_\mu e_a
=-\frac{1}{4}e^a\Upsilon_\mu e_a\in\Lambda_2\top_1.
\label{B-def}
\end{equation}

\theorem 4. Under the Lorentz rotation of tetrad (\ref{rot:tetrad})
the tensor $B_\mu$ transforms as a connection
$$
B_\mu\to\check B_\mu=S^{-1}B_\mu S-S^{-1}\Upsilon_\mu S.
$$

\proof. We have
\begin{eqnarray*}
&&-4\check B_\mu=\check e^a\Upsilon_\mu\check e_a=S^{-1}e^a S\Upsilon_\mu(S^{-1}e_a S)\\
&&=S^{-1}e^a S(\Upsilon_\mu S^{-1})e_a S+S^{-1}e^a \Upsilon_\mu(e_a) S+
S^{-1}e^a e_a\Upsilon_\mu S\\
&&=-4S^{-1}B_\mu S+4S^{-1}\Upsilon_\mu S+
S^{-1}e^a S(\Upsilon_\mu S^{-1})e_a S.
\end{eqnarray*}
Here we use the formula $e^a e_a=4$ from Theorem 3. It can be checked that
$S\Upsilon_\mu S^{-1}\in\Lambda_2\top_1$. Consequently from Theorem 3 we have
$$
e^a S\Upsilon_\mu S^{-1} e_a S=0.
$$
These completes the proof.
\medskip

From the formula (\ref{B-def}) it is easily shown that
$$
\Upsilon_\mu e^a=[B_\mu,e^a],\quad
\Upsilon_\mu e_a=[B_\mu,e_a].
$$
Hence, if we take the operators
$$
\D_\mu=\Upsilon_\mu-[B_\mu,\,\cdot\,],
$$
then
$$
\D_\mu e^a=0,\quad \D_\mu e_a=0.
$$

\theorem 5. The operators $\D_\mu$ satisfy the Leibniz rule
$$
\D_\mu(UV)=(\D_\mu U)V+U\D_\mu V\quad\hbox{for}\quad
U\in\Lambda\top^r_s,\,V\in\Lambda\top^p_q
$$
and
$$
\D_\mu\D_\nu-\D_\nu\D_\mu=0.
$$
\par
\proof\,  is by direct calculations.
\medskip

Note that {\em the volume form}
$$
\ell=e^0\wedge e^1\wedge e^2\wedge e^3=\sqrt{-g}\,dx^0\ww dx^3\in\Lambda_4
$$
is constant with respect to these operators, i.e.,
$$
\D_\mu\ell=0.
$$

We have the following consequences from the theorems 4,5. Under the
Lorentz rotation of tetrad (\ref{rot:tetrad}) $B_\mu,\D_\mu$ transform as follows
\begin{eqnarray}
B_\mu&\to&\check B_\mu=B_\mu-S^{-1}\D_\mu S,\label{B:transform}\\
\D_\mu&\to&\check\D_\mu=\D_\mu+[S^{-1}\D_\mu S,\,\cdot\,].
\nonumber
\end{eqnarray}

Let us denote 1-form
$$
H:=e^0.
$$
We define the operator of {\em Hermitian conjugation} of tensors
$\dagger\,:\,\Lambda\top^p_q\to\Lambda\top^p_q$
$$
U^\dagger=H U^* H \quad\hbox{for}\quad U\in\Lambda\top^p_q.
$$
Evidently,
$$
(UV)^\dagger=V^\dagger U^\dagger,\quad U^{\dagger\dagger}=U,
\quad i^\dagger=-i.
$$
We shall see in Section 10 that this operator is connected to the
operator of Hermitian conjugation of matrices.

We say that a tensor $U\in\Lambda\top^p_q$ is {\em Hermitian} if
$U^\dagger=U$ and {\em anti-Hermitian} if $U^\dagger=-U$. Every tensor
$U$ can be decomposed into Hermitian and anti-Hermitian parts
$$
U=\frac{1}{2}(U+U^\dagger)+\frac{1}{2}(U-U^\dagger).
$$

Note that all discussed constructions, which were defined in this paper
for tensors from $\Lambda\top^p_q$, are also valid for complex valued
tensors from $\Lambda^\C\top^p_q$.

Now we may define the operation
$$
(\cdot,\cdot)\,:\,\Lambda^\C\times\Lambda^\C\to\Lambda^\C_0
$$
by the formula
$$
(U,V)=\Tr(U^\dagger V).
$$
This operation has all the properties of Hermitian scalar product at
every point $x\in X$
\begin{eqnarray*}
&&\alpha(U,V)=(\bar\alpha U,V)=(U,\alpha V),\\
&&(U,V)=\overline{(V,U)},\quad (U+W,V)=(U,V)+(W,V),\\
&&(U,U)>0\quad\hbox{for}\quad U\neq0,
\end{eqnarray*}
where $U,V,W\in\Lambda^\C$, $\alpha\in\Lambda^\C_0$, and a bar means
complex conjugation. The operation $(\cdot,\cdot)$ converts $\Lambda^\C$
into the unitary space at every point $x\in X$.

Let us denote by $T_0,\ldots T_{15}$ the following differential forms:
\begin{equation}
i,ie^0,e^1,e^2,e^3,ie^{01},ie^{02},ie^{03},e^{12},e^{13},e^{23},
e^{012},e^{013},e^{023},ie^{123},e^{0123}.
\label{basis:ah}
\end{equation}
which form an orthonormal basis of $\Lambda^\C$
$$
(T_\k,T^\n)=\delta_\k^\n,\quad {\textsc k},{\textsc n}=0,\ldots15
$$
and
$$
T_\k=-T_\k^\dagger,\quad \D_\mu T_\k=0,
$$
where $T^\n=T_\n$. This basis is said to be {\em the anti-Hermitian basis
of $\Lambda^\C$}.


A differential form $t\in\Lambda^\C$ such that
\begin{equation}
t^2=t,\quad \D_\mu t=0,\quad t^\dagger=t
\label{t}
\end{equation}
is called {\em an idempotent}. We suppose that under a Lorentz rotation
of tetrad $e^a\to\check e^a=S^{-1}e^a S$, $S\in\Spin(\W)$ an idempotent
$t$ transforms as
$$
t\to\check t=S^{-1}tS.
$$
In this case
\begin{eqnarray*}
&&\check t^2=\check t,\\
&&\check t^\dagger=\check H\check t^*\check H=\check t,\\
&&\check\D_\mu\check t=0.
\end{eqnarray*}

We may consider {\em the left ideal} generated by the idempotent $t$
\begin{equation}
\I(t)=\{Ut\,:\,U\in\Lambda^\C\}\subseteq\Lambda^\C.
\label{ideal}
\end{equation}

Let us define the set of differential forms
$$
L(t)=\{U\in\I(t)\,:\,U^\dagger=-U,\,[U,t]=0\}.
$$
This set is closed with respect to the commutator (if $U,V\in L(t)$,
then $[U,V]\in L(t)$) and can be considered as a real Lie algebra.
With the aid of the real Lie algebra $L(t)$ we define the corresponding
Lie group
$$
G(t)=\{\exp(U)\,:\,U\in L(t)\}.
$$
In Section 10 we consider $t,\I(t),L(t),G(t)$ in details.


Finally let us summarize properties of the operators $\D_\mu$
\begin{eqnarray*}
&&\D_\mu(UV)=(\D_\mu U)V+U\D_\mu V,\\
&&\D_\mu\D_\nu-\D_\nu\D_\mu=0,\\
&&\D_\mu(U+V)=\D_\mu U+\D_\mu V,\\
&&\D_\mu e^a=0,\,\,\D_\mu e_a=0,\\
&&\D_\mu\ell=0,\\
&&\D_\mu(U^*)=(\D_\mu U)^*,\\
&&\D_\mu(U^\dagger)=(\D_\mu U)^\dagger,\\
&&\D_\mu(\Tr(U))=\partial_\mu(\Tr(U))=\Tr(\D_\mu U),\\
&&\D_\mu(U,V)=\partial_\mu(U,V)=(\D_\mu U,V)+(U,\D_\mu V).
\end{eqnarray*}
Under a change of coordinates $(x)\to(\acute x)$ operators $\D_\mu$ transform as components of a covector, i.e.,
$$
\D_\mu\to\acute\D_\mu=\frac{\partial x^\nu}{\partial\acute x^\mu}\D_\nu.
$$
Under a Lorentz rotation
of tetrad $e^a\to\check e^a=S^{-1}e^a S$ ($S\in\Spin(\W)$)
operators $\D_\mu$ transform as
$\D_\mu\to\check\D_\mu=\D_\mu+[S^{-1}\D_\mu S,\,\cdot\,]$.


\section{Dirac-type tensor equations. A general case.}
We begin with the following equation in $\W$ (a tetrad $e^a$ is given and, consequently, the tensor
$B_\mu\in\Lambda_2\top_1$ and the operators $\D_\mu$ are defined):
\begin{equation}
dx^\mu(\D_\mu\Phi+\Phi A_\mu+B_\mu\Phi)+im\Phi=0,
\label{init:eq}
\end{equation}
where $\Phi\in\Lambda^\C$, $i=\sqrt{-1}$, $m$ is a given real constant, and $A_\mu\in\Lambda^\C\top_1$ is such
that $A_\mu{}^\dagger=-A_\mu$. We consider the differential form $\Phi$ as unknown (16 complex valued components)
and $A_\mu$ as known. Writing  eq. (\ref{init:eq}) as a system of
equations for components of $\Phi$, we see that the number of equations is equal to the number of
unknown values.
\medskip

An equation is said to be
{\em a tensor equation} if all values in it are tensors and all operations in it
take tensors to tensors.
\medskip

In eq. (\ref{init:eq}) we have $dx^\mu\D_\mu\Phi,dx^\mu\Phi A_\mu,dx^\mu B_\mu\Phi\in\Lambda^\C$.
Hence eq. (\ref{init:eq}) is a tensor equation.

Let $t\in\Lambda^\C$ be an idempotent and $A_\mu$ be a tensor from $L(t)\top_1$. Then we may
consider the equation in $\W$
\begin{equation}
(dx^\mu(\D_\mu\Phi+\Phi A_\mu+B_\mu\Phi)+im\Phi)t=0,
\label{init2:eq}
\end{equation}
From the identities $\D_\mu t=0$, $[A_\mu,t]=0$ it follows that eq. (\ref{init2:eq})
can be written as the equation for $\Psi=\Phi t\in\I(t)$
\begin{equation}
dx^\mu(\D_\mu\Psi+\Psi A_\mu+B_\mu\Psi)+im\Psi=0,
\label{Dirac:type}
\end{equation}
where the idempotent $t$, the real constant $m$, and the tensor $A_\mu\in L(t)\top_1$
are considered as known. The differential form $\Psi\in\I(t)$ is considered as unknown.

In Section 10 we shall see that there are four types of idempotents $t$. Consequently, there
are four types of equations (\ref{Dirac:type}). These equations are called
{\em Dirac-type tensor equations}. A connection of eqs. (\ref{Dirac:type}) with the
Dirac equation will be discussed in Section 9.

Denoting $\alpha^\mu=H dx^\mu\in(\Lambda_0\oplus\Lambda_2)\top^1$, we see that
$(\alpha^\mu)^\dagger=\alpha^\mu$.

\theorem 6. If $\Psi\in\I(t)$ satisfies eq. (\ref{Dirac:type}), then the tensor
$$
J^\mu=i\Psi^\dagger\alpha^\mu\Psi
$$
satisfies the equality
\begin{equation}
\frac{1}{\sqrt{-g}}\D_\mu(\sqrt{-g}\,J^\mu)-[A_\mu,J^\mu]=0,
\label{conserv:law}
\end{equation}
which is called {\em a (non-Abelian) charge conservation law}.
\par

Note that $\Psi=\Psi t$ and $J^\mu=t J^\mu t$. Therefore,
$$
[J^\mu,t]=[t J^\mu t,t]=0,
$$
i.e., $J^\mu\in L(t)\top^1$.

\proof\, of Theorem 6. Let us multiply eq. (\ref{init2:eq}) from the left by $H$ and
denote the left-hand side of resulting equation by
\begin{equation}
Q=\alpha^\mu(\D_\mu\Psi+\Psi A_\mu+B_\mu\Psi)+imH\Psi.
\label{Q}
\end{equation}
Then
$$
Q^\dagger=(\D_\mu\Psi^\dagger-A_\mu\Psi^\dagger+\Psi^\dagger B_\mu^\dagger)\alpha^\mu-
im\Psi^\dagger H.
$$
Consider the expression
\begin{eqnarray}
i(\Psi^\dagger Q+Q^\dagger\Psi)&=&
i(\Psi^\dagger\alpha^\mu\D_\mu\Psi+
\D_\mu\Psi^\dagger\alpha^\mu\Psi+\Psi^\dagger(\D_\mu\alpha^\mu)\Psi)\nonumber
-[A_\mu,i\Psi^\dagger\alpha^\mu\Psi]\\
&&+i\Psi^\dagger(-D_\mu\alpha^\mu+\alpha^\mu B_\mu+B_\mu^\dagger\alpha^\mu)\Psi\label{tmp}\\
&=&\frac{1}{\sqrt{-g}}\D_\mu(\sqrt{-g}\,J^\mu)-[A_\mu,J^\mu].\nonumber
\end{eqnarray}
Here we use the formulae
$$
\D_\mu\alpha^\mu=-\Gamma_{\mu\nu}{}^\mu \alpha^\nu+\alpha^\mu B_\mu+B_\mu^\dagger\alpha^\mu,
$$
$$
\D_\mu J^\mu+\Gamma_{\mu\nu}{}^\mu
J^\nu=\frac{1}{\sqrt{-g}}\D_\mu(\sqrt{-g}\,J^\mu).
$$
which can be easily checked.
In (\ref{tmp}) equality $Q=0$ leads to the equality
(\ref{conserv:law}). These completes the proof.
\medskip

Now we may write eq. (\ref{Dirac:type}) together with {\em Yang-Mills equations}
(\ref{YM1},\ref{YM2})
\begin{eqnarray}
&&dx^\mu(\D_\mu\Psi+\Psi A_\mu+B_\mu\Psi)+im\Psi=0, \label{Dirac:type1}\\
&&\D_\mu A_\nu-\D_\nu A_\mu-[A_\mu,A_\nu]=F_{\mu\nu},\label{YM1}\\
&&\frac{1}{\sqrt{-g}}\D_\mu(\sqrt{-g}\,F^{\mu\nu})-[A_\mu,F^{\mu\nu}]=J^\nu,\label{YM2}\\
&&J^\nu=i\Psi^\dagger\alpha^\nu\Psi,\label{J}
\end{eqnarray}
where $\Psi\in\I(t)$, $A_\mu\in L(t)\top_1$, $F_{\mu\nu}\in L(t)\top_2$, $J^\nu\in L(t)\top^1$.
In this system of equations we consider $\Psi,A_\mu,F_{\mu\nu},J^\nu$ as unknown values and
$m,t$ as known values.

\theorem 7. Let us denote the left-hand side of eq. (\ref{YM2}) by $R^\nu$
$$
R^\nu:=\frac{1}{\sqrt{-g}}\D_\mu(\sqrt{-g}\,F^{\mu\nu})-[A_\mu,F^{\mu\nu}],
$$
where $F_{\mu\nu}$ satisfy (\ref{YM1}). Then
$$
\frac{1}{\sqrt{-g}}\D_\mu(\sqrt{-g}\,R^{\mu})-[A_\mu,R^{\mu}]=0.
$$
\par

The proof is by direct calculations.
\medskip

This theorem means that eq. (\ref{YM2}) is consistent with the charge conservation law
(\ref{conserv:law}).


\section{Unitary and Spin gauge symmetries.}
\theorem 8. Let $\Psi,A_\mu,F_{\mu\nu},J^\nu$ satisfy eqs. (\ref{Dirac:type1}-\ref{J})
with a given idempotent $t$ and constant $m$. And let $U\in G(t)$, where the Lie group
$G(t)$ is defined in Section 6. Then the following values with tilde:
$$
\tilde\Psi=\Psi U,\quad\tilde A_\mu=U^{-1}A_\mu U-U^{-1}\D_\mu U,\quad
\tilde F_{\mu\nu}=U^{-1}F_{\mu\nu}U,
$$
$$
\tilde J^\nu=U^{-1}J^\nu U,\quad
\{\tilde t,\tilde B_\mu,\tilde\D_\mu\}=
\{t,B_\mu,\D_\mu\}
$$
satisfy the same equations (\ref{Dirac:type1}-\ref{J}).
\par

The proof is straightforward.
\medskip

This theorem means that eqs. (\ref{Dirac:type1}-\ref{J}) are invariant under gauge
transformations with the symmetry Lie group $G(t)$.

\theorem 9.  Let $\Psi,A_\mu,F_{\mu\nu},J^\nu$ satisfy eqs. (\ref{Dirac:type1}-\ref{J})
with a given idempotent $t$ and constant $m$. And let $S\in\Spin(\W)$.
Then the following values with check:
\begin{equation}
\check\Psi=\Psi S,\quad\check A_\mu=S^{-1}A_\mu S,\quad
\check F_{\mu\nu}=S^{-1}F_{\mu\nu}S,\quad\check J^\nu=S^{-1}J^\nu S,
\label{theorem9}
\end{equation}
$$
\check t=S^{-1}tS,\quad \check B_\mu=B_\mu-S^{-1}\D_\mu S,
\quad \check\D_\mu=\D_\mu+[S^{-1}\D_\mu,\,\cdot\,]
$$
satisfy the same equations (\ref{Dirac:type1}-\ref{J}).
\par

The proof is by direct calculations.
\medskip

Note that for values with check the operation of Hermitian conjugation is defined
by $\check U^\dagger=\check H\check U^*\check H$, where $\check H=S^{-1}H S$.
\medskip

This theorem means that eqs. (\ref{Dirac:type1}-\ref{J})
are invariant under gauge transformations with the symmetry Lie group $\Spin(\W)$.

Eqs. (\ref{Dirac:type1}-\ref{J}) can be derived from the following Lagrangian:
\begin{equation}
\L=\frac{1}{4}i\sqrt{-g}\,\Tr(\Psi^\dagger H Q-Q^\dagger H\Psi)+
C\frac{1}{4}\sqrt{-g}\,Tr(\frac{1}{8}F_{\mu\nu}F^{\mu\nu}),
\label{lagr}
\end{equation}
where $Q$ is from (\ref{Q}), $F_{\mu\nu}=\D_\mu A_\nu-\D_\nu A_\mu-[A_\mu,A_\nu]
\in L(t)\top_2$, and $C$ is a real constant.
If we have a basis $\{t_1,\ldots,t_D\}$ of $\I(t)$ and a basis
$\{\tau_1,\ldots,\tau_d\}$ of
$L(t)$ that satisfy (\ref{basis}),(\ref{tau}),
then we may substitute $\Psi=\psi^\k t_\k$,
$A_\mu=a_\mu^n \tau_n$ into the Lagrangian $\L$.
Variating $\L$ with respect to $\psi^\k$
and $a_\mu^n$, we arrive at eqs. (\ref{Dirac:type1}-\ref{J}). We discuss bases of
$\I(t)$ and $L(t)$ in Section 10.

In \cite{nona2} we discuss a gravitational term for the Lagrangian $\L$.


\section{A connection between the Dirac-type tensor equation and the Dirac equation.}
Let $t\in\Lambda^\C$ be an idempotent with the properties (\ref{t}) and $\I(t)$ be the
left ideal (\ref{ideal}) of complex dimension $D$. In the next section we shall see that
$D$ may take one of four possible values $D=4,8,12,16$. We use an orthonormal basis
$t_{\k}=t^{\k}$, ${\sc k}=1,\ldots,D$ of $\I(t)$ such that
\begin{equation}
\D_\mu t_{\k}=0,\quad (t_{\n},t^{\k})=\delta^{\k}_{\n}.
\label{basis}
\end{equation}
{\sc Small Caps} font indices run from $1$ to $D$. Consider a linear operator $|\rangle $ that
maps $\I(t)$ to $\C^D$. If
$$
\Omega=\omega^{\k}t_{\k}\in\I(t),
$$
then
$$
|\Omega\rangle =(\omega^1\ldots\omega^D)^{\rm T}.
$$
In particular,
$$
|t_{\k}\rangle =(0\,\ldots\,1\,\ldots\,0)^{\rm T}
$$
with only $1$ on the $k$-th place of the column.

By $M^\C(D)\top_p^q$ denote the set of type $(p,q)$ tensors with values in
$D\!\times\!D$ complex matrices. Let $\gamma\,:\,\Lambda^\C\top^q_p\to M^\C(D)\top^q_p$
be a map such that for $U=(U^{\nu_1\ldots\nu_q}_{\mu_1\ldots\mu_p})\in\Lambda^\C\top_p^q$
\begin{equation}
Ut_{\n}=\gamma(U)^{\k}_{\n}t_{\k}.
\label{gamma:map}
\end{equation}
Hence,
$$
\gamma(U)^{\k}_{\n}=(t^{\k},Ut_{\n}),
$$
where $\gamma(U)^{\k}_{\n}$ are elements of the matrix $\gamma(U)$ (an upper
index enumerates rows and a lower index enumerates columns).

If $U\in\Lambda^\C$ and $\Omega\in\I(t)$, then
$$
U\Omega=U\omega^{\n}t_{\n}=\omega^{\n}\gamma(U)^{\k}_{\n}t_{\k}.
$$
That means
$$
|U\Omega\rangle =\gamma(U)|\Omega\rangle .
$$
If $U,V\in\Lambda^\C$, $\Omega\in\I(t)$, then
$$
|UV\Omega\rangle =\gamma(U)\gamma(V)|\Omega\rangle =\gamma(UV)|\Omega\rangle .
$$
Consequently,
$$
\gamma(UV)=\gamma(U)\gamma(V),
$$
i.e., $\gamma$ is a matrix representation of $\Lambda^\C$. For example, if we take
$dx^\mu=\delta^\mu_\nu dx^\nu\in\Lambda_1\top^1$, then we get
$$
dx^\mu t_{\n}=\gamma(dx^\mu)^{\k}_{\n}t_{\k}.
$$
Denoting $\gamma^\mu=\gamma(dx^\mu)$, we see that the equality
$dx^\mu dx^\nu+dx^\nu dx^\mu=2g^{\mu\nu}$ leads to the equality
$$
\gamma^\mu\gamma^\nu+\gamma^\nu\gamma^\mu=2g^{\mu\nu}{\bf 1},
$$
where ${\bf 1}$ is the identity matrix of dimension $D$.

Let us take the set of differential forms
$$
\K(t)=\{V\in\Lambda^\C\,:\,[V,t]=0\},
$$
which can be considered as an algebra (at any point $x\in X$). Now we define
a map
$$
\theta\,:\,\K(t)\top^q_p\to M^\C(D)\top^q_p
$$
such that for
$V=(V^{\nu_1\ldots\nu_q}_{\mu_1\ldots\mu_p})\in\K(t)\top^q_p$
$$
t_{\n}V=\theta(V)^{\k}_{\n}t_{\k}.
$$
Therefore,
$$
\theta(V)^{\k}_{\n}=(t^{\k},t_{\n}V).
$$
If $V\in\K(t)$ and $\Omega\in\I(t)$, then
$$
\Omega V=\omega^{\n}t_{\n}V=\omega^{\n}\theta^{\k}_{\n}t_{\k}.
$$
Than means
$$
|\Omega V\rangle =\theta(V)|\Omega\rangle .
$$
If $U,V\in\K(t)$, $\Omega\in\I(t)$, then
$$
|\Omega UV\rangle =\theta(V)\theta(U)|\Omega\rangle =\theta(UV)|\Omega\rangle .
$$
Consequently,
$$
\theta(UV)=\theta(V)\theta(U).
$$

If $U\in\Lambda^\C$, $V\in\K(t)$, $\Omega\in\I(t)$, then
$U\Omega\in\I(t)$, $\Omega V\in\I(t)$ and
\begin{eqnarray*}
|U\Omega V\rangle &=&\gamma(U)|\Omega V\rangle =\gamma(U)\theta(V)|\Omega\rangle ,\\
|U\Omega V\rangle &=&\theta(V)|U\Omega\rangle =\theta(V)\gamma(U)|\Omega\rangle .
\end{eqnarray*}
Consequently,
$$
[\gamma(U),\theta(V)]=0.
$$

Denoting the left-hand side of eq. (\ref{Dirac:type}) by
\begin{equation}
\Omega=dx^\mu(\D_\mu\Psi+\Psi A_\mu+B_\mu\Psi)+im\Psi
\label{Omega}
\end{equation}
and using formulas
\begin{eqnarray*}
&&\psi:=|\Psi\rangle ,\\
&&|\D_\mu\Psi\rangle =|\D_\mu(t_{\k}\psi^{\k})\rangle =|t_{\k}\partial_\mu\psi^{\k}\rangle=
\partial_\mu\psi,\\
&&|dx^\mu\Psi A_\mu\rangle =\gamma^\mu|\Psi A_\mu\rangle =\gamma^\mu\theta(A_\mu)\psi,\\
&&|dx^\mu B_\mu\Psi\rangle =\gamma^\mu|B_\mu\Psi\rangle =\gamma^\mu\gamma(B_\mu)\psi,
\end{eqnarray*}
we see that
\begin{equation}
|\Omega\rangle =\gamma^\mu(\partial_\mu+\theta(A_\mu)+\gamma(B_\mu))\psi+im\psi.
\label{Omega:column}
\end{equation}
Note that
$$
B_\mu=\frac{1}{2}b_{\mu ab}e^a\wedge e^b=\frac{1}{4}b_{\mu ab}(e^a e^b-e^b e^a).
$$
Therefore,
$$
\gamma(B_\mu)=\frac{1}{4}b_{\mu ab}[\gamma^a,\gamma^b],
$$
where $\gamma^a=\gamma(e^a)$ and the equalities $e^a e^b+e^b e^a=2\eta^{ab}$
leads to the equalities
$$
\gamma^a\gamma^b+\gamma^b\gamma^a=2\eta^{ab}{\bf 1}.
$$
If we take $V\in G(t)$, then
\begin{eqnarray*}
\Omega V&=&(dx^\mu(\D_\mu\Psi+\Psi A_\mu+B_\mu\Psi)+im\Psi)V\\
&=&dx^\mu(\D_\mu(\Psi V)+\Psi V(V^{-1}A_\mu V-V^{-1}\D_\mu V)+B_\mu(\Psi V))+im(\Psi V)
\end{eqnarray*}
and
\begin{eqnarray*}
|\Omega V\rangle  &=&\theta(V)|\Omega\rangle =P(\gamma^\mu(\partial_\mu+\theta(A_\mu)+\gamma(B_\mu))\psi+im\psi)\\
&=&\gamma^\mu(\partial_\mu+(P\theta(A_\mu)P^{-1}-(\partial_\mu P)P^{-1})+\gamma(B_\mu))(P\psi)+imP\psi,
\end{eqnarray*}
where $P=\theta(V)$ and we use formulae $[P,\gamma^\mu]=0$, $[P,\gamma(B_\mu)]=0$. Thus the invariance
of eq. (\ref{Dirac:type}) under the gauge transformation ($V\in G(t)$)
$$
\Psi\to\Psi V,\quad A_\mu\to V^{-1}A_\mu V-V^{-1}\D_\mu V
$$
leads to  the invariance of the equation
\begin{equation}
\gamma^\mu(\partial_\mu+\theta(A_\mu)+\gamma(B_\mu))\psi+im\psi=0
\label{psi}
\end{equation}
under the gauge transformation ($P=\theta(V)$)
$$
\psi\to P\psi,\quad \theta(A_\mu)\to P\theta(A_\mu)P^{-1}-(\partial_\mu P)P^{-1}.
$$
Let $S\in\Spin(\W)$. Consider the Lorentz rotation of the tetrad
$e^a\to\check e^a=S^{-1}e^a S$, which leads to the transformation
$$
t\to\check t=S^{-1}tS,\quad \I(t)\to\I(\check t),\quad t_{\n}\to\check t_{\n}=S^{-1}t_{\n}S.
$$
Let us define a map $|\check\rangle\,:\,\I(\check t)\to\C^D$.
If $\Phi=\phi^{\k}\check t_{\k}\in\I(\check t)$, then
$$
|\Phi\check\rangle=(\phi^1\ldots\phi^D)^{\rm T}.
$$
Evidently for $\Omega\in\I(t)$
$$
|S^{-1}\Omega S\check\rangle=|\Omega\rangle.
$$
Also we may define a map
$\check\gamma\,:\,\Lambda^\C\top_p^q\to M^\C(D)\top^q_p$ such that
$$
U\check t_{\n}=\check\gamma(U)^{\k}_{\n}\check t_{\k}
$$
and
$$
\check\gamma(UV)=\check\gamma(U)\check\gamma(V).
$$
It is easily seen that
$$
\check\gamma(U)=\gamma(S)\gamma(U)\gamma(S^{-1}).
$$
In particular,
$$
\check\gamma(S)=\gamma(S).
$$
For $\Omega\in\I(t)$ we have
$$
|\Omega S\check\rangle=|S(S^{-1}\Omega S)\check\rangle=
\check\gamma(S)|S^{-1}\Omega S\check\rangle=\gamma(S)|\Omega\rangle.
$$
If we apply these identities to
\begin{eqnarray*}
\Omega S&=&(dx^\mu(\D_\mu\Psi+\Psi A_\mu+B_\mu\Psi)+im\Psi)S\\
&=&dx^\mu(\check\D_\mu\check\Psi+\check\Psi\check A_\mu+\check
B_\mu\check\Psi)+im\check\Psi,
\end{eqnarray*}
then we get
\begin{eqnarray*}
|\Omega S\check\rangle&=&\gamma(S)|\Omega\rangle=
R(\gamma^\mu(\partial_\mu\psi+\theta(A_\mu)\psi+\gamma(B_\mu)\psi)+im\psi)\\
&=&R\gamma^\mu R^{-1}(\partial_\mu(R\psi)+\theta(A_\mu)R\psi+
(R\gamma(B_\mu)R^{-1}-(\partial_\mu R)R^{-1})R\psi) +imR\psi,
\end{eqnarray*}
where $R=\gamma(S)$. Note that $[R,\theta(A_\mu)]=0$.

This implies that the invariance of eq. (\ref{Dirac:type}) under the
gauge transformation (\ref{theorem9})
leads to the invariance of eq. (\ref{psi}) under the gauge
transformation ($R=\gamma(S)$)
$$
\psi\to R\psi,\quad \theta(A_\mu)\to\theta(A_\mu)\quad
\gamma^\mu\to R\gamma^\mu R^{-1},
$$
$$
\gamma(B_\mu)\to R\gamma(B_\mu)R^{-1}-(\partial_\mu R)R^{-1}.
$$

Now consider a transformation of eqs. (\ref{Dirac:type}),(\ref{psi}) under a
change of coordinates $(x)\to(\acute x)$. Coordinates $(\acute x)$ we denote
with the aid of primed indices $x^{\mu^\prime}$. In coordinates $(\acute x)$ eq.
(\ref{Dirac:type}) has the form
$$
\acute\Omega\equiv dx^{\mu^\prime}(\D_{\mu^\prime}\Psi+\Psi A_{\mu^\prime}+
B_{\mu^\prime}\Psi)+im\Psi=0,
$$
where
$$
\D_{\mu^\prime}=\frac{\partial x^{\nu}}{\partial x^{\mu^\prime}}\D_\nu,\quad
dx^{\mu^\prime}=\frac{\partial x^{\mu^\prime}}{\partial x^\nu}dx^\nu,\quad
A_{\mu^\prime}=\frac{\partial x^{\nu}}{\partial x^{\mu^\prime}}A_\nu,\quad
B_{\mu^\prime}=\frac{\partial x^{\nu}}{\partial x^{\mu^\prime}}B_\nu,
$$
and eq. (\ref{psi}) has the form
$$
|\acute\Omega\rangle=
\gamma^{\mu^\prime}(\partial_{\mu^\prime}+\theta(A_{\mu^\prime})+
\gamma(B_{\mu^\prime}))\psi+im\psi=0,
$$
where
$$
\partial_{\mu^\prime}=\frac{\partial}{\partial x^{\mu^\prime}},\quad
\gamma^{\mu^\prime}=
\frac{\partial x^{\mu^\prime}}{\partial x^\nu}\gamma^\nu.
$$
Note that
$$
\gamma^\mu=\gamma(dx^\mu)=\gamma^a e^\mu{}_a,\quad
\gamma^a=\gamma(e^a)
$$
and
$$
\gamma^{\mu^\prime}=\gamma^a e^{\mu^\prime}{}_a=
\gamma^a\frac{\partial x^{\mu^\prime}}{\partial x^\nu}e^\nu{}_a.
$$

Finally, let us note that in the Lagrangian (\ref{lagr})
$$
\frac{1}{4}\sqrt{-g}\,\Tr(i(\Psi^\dagger H\Omega-\Omega^\dagger
H\Psi))=
\frac{1}{4}\sqrt{-g}\,i(\langle\Psi|\gamma^0|\Omega\rangle-
\langle\Omega|\gamma^0|\Psi\rangle),
$$
where $\langle\Psi|=|\Psi\rangle^\dagger$.


\section{Idempotents and  bases of left ideals.}
Let us take the idempotent
$$
t_{(1)}=\frac{1}{4}(1+e^0)(1+i e^{12})=
\frac{1}{4}(1+e^0+i e^{12}+i e^{012}),
$$
which satisfies conditions (\ref{t}), and consider the left ideal $\I(t_{(1)})$.
It can be shown that the complex dimension of $\I(t_{(1)})$ is equal
to four (this $\I(t_{(1)})$ is {\em a minimal left ideal} of
$\Lambda^\C$ and $t_{(1)}$ is {\em a primitive idempotent})
and the following differential forms
\begin{eqnarray*}
t^1&=&2 t_{(1)}=\frac{1}{2}(1+e^0+i e^{12}+i e^{012}),\\
t^2&=&-2e^{13} t_{(1)}=\frac{1}{2}(-e^{13}+ie^{23}-e^{013}+ie^{023}),\\
t^3&=&2e^{03} t_{(1)}=\frac{1}{2}(-e^3+e^{03}-ie^{123}+ie^{0123}),\\
t^4&=&2e^{01} t_{(1)}=\frac{1}{2}(-e^1+ie^2+e^{01}-ie^{02})
\end{eqnarray*}
can be taken as basis forms of $\I(t_{(1)})$, which satisfy (\ref{basis}).
This basis, according to the formula (\ref{gamma:map}), defines the
matrix representation of $\Lambda^\C$ ($\gamma_{(1)}$ is a one-to-one map)
$$
\gamma_{(1)}\,:\,\Lambda^\C\top^q_p\to M^\C\top^q_p.
$$

In particular we get matrices $\gamma_{(1)}^\mu=\gamma_{(1)}(dx^\mu)$
identical to (\ref{gamma:matrices}). Denote
$$
\underline U:=\gamma_{(1)}(U)\for U\in\Lambda^\C\top^q_p.
$$
Let $Y^\n_\k\in\Lambda^\C$, ($\textsc k,\textsc n=1,2,3,4$) be differential forms such
that $\underline Y^\n_\k$ are $4\!\times\!4$-matrices with only nonzero element
that equal to $1$ on the intersection of ${\textsc n}$-th row and
${\textsc k}$-th column. We can calculate that
\begin{eqnarray*}
Y^1_1&=& (1+ e^{0}+ e^{012}i + e^{12}i)/4,\\
Y^1_2&=& (e^{013}+ e^{13}+ e^{023}i + e^{23}i)/4,\\
Y^1_3&=& (e^{03}+ e^{3}+ e^{0123}i + e^{123}i)/4,\\
Y^1_4 &=& (e^{01}+ e^{1}+ e^{02}i + e^{2}i)/4,\\
Y^2_1&=& (-e^{013}- e^{13}+ e^{023}i + e^{23}i)/4,\\
Y^2_2&=& (1+ e^{0}- e^{012}i - e^{12}i)/4,\\
Y^2_3&=& (e^{01}+ e^{1}- e^{02}i - e^{2}i)/4,\\
Y^2_4 &=& (-e^{03}- e^{3}+ e^{0123}i + e^{123}i)/4,\\
Y^3_1&=& (e^{03}- e^{3}+ e^{0123}i - e^{123}i)/4,\\
Y^3_2&=& (e^{01}- e^{1}+ e^{02}i - e^{2}i)/4,\\
Y^3_3&=& (1- e^{0}- e^{012}i + e^{12}i)/4,\\
Y^3_4 &=& (-e^{013}+ e^{13}- e^{023}i + e^{23}i)/4,\\
Y^4_1&=& (e^{01}- e^{1}- e^{02}i + e^{2}i)/4,\\
Y^4_2&=& (-e^{03}+ e^{3}+ e^{0123}i - e^{123}i)/4,\\
Y^4_3&=& (e^{013}- e^{13}- e^{023}i + e^{23}i)/4,\\
Y^4_4 &=& (1- e^{0}+ e^{012}i - e^{12}i)/4.
\end{eqnarray*}
We see that
$$
t_{(1)}=Y^1_1,\quad \underline t_{(1)}=\diag(1,0,0,0)
$$
and
$$
t^1=2Y^1_1,\quad t^2=2Y^2_1,\quad t^3=2Y^3_1,\quad t^4=2Y^4_1.
$$
Now we may define the idempotents
\begin{eqnarray*}
t_{(2)}&=&Y^1_1+Y^2_2=\frac{1}{2}(1+e^0),\\
t_{(3)}&=&Y^1_1+Y^2_2+Y^3_3=\frac{1}{4}(3+e^0+ie^{12}-ie^{012}),\\
t_{(4)}&=&Y^1_1+Y^2_2+Y^3_3+Y^4_4=1
\end{eqnarray*}
such that
$$
\underline t_{(2)}=\diag(1,1,0,0),
\quad \underline t_{(3)}=\diag(1,1,1,0),
\quad \underline t_{(4)}=\diag(1,1,1,1).
$$
Also we can take the following differential forms $t^1,\ldots, t^{16}$:
$$
t^{4(\n-1)+\k}=2Y^\k_\n,\quad \textsc k,\textsc n=1,2,3,4,
$$
which satisfy conditions $\D_\mu t^\k=0$, $(t_\k,t^\n)=\delta^\n_\k$.
Evidently, $\{t_1,\ldots,t_8\}$ is a basis of $\I(t_{(2)})$,
$\{t_1,\ldots,t_{12}\}$ is a basis of $\I(t_{(3)})$, and
$\{t_1,\ldots,t_{16}\}$ is a basis of $\I(t_{(4)})=\Lambda^\C$. In
accordance with the formula (\ref{gamma:map}), these bases define the
maps
$$
\gamma_{(k)}\,:\,\Lambda^\C\top^q_p\to M^\C(4k)\top^q_p,\quad
k=1,2,3,4
$$
such that
$$
\gamma_{(k)}(UV)=\gamma_{(k)}(U)\gamma_{(k)}(V)\for
U\in\Lambda^\C\top^q_p,\,V\in\Lambda^\C\top_r^s.
$$
Also the maps $\gamma_{(k)}$ have the important property
$$
\gamma_{(k)}(U^\dagger)=(\gamma_{(k)}(U))^\dagger\for
U\in\Lambda^\C\top^q_p,
$$
where $U^\dagger=H U^* H$ is the Hermitian-conjugated differential
form and $(\gamma_{(k)(U)})^\dagger$ is the Hermitian conjugated matrix
(transposed matrix with complex conjugated elements).
Consider the set of differential forms
$$
\K_0(t)=\{U\in\I(t)\,:\,[U,t]=0\}=\K(t)\cap\I(t)
$$
and the corresponding set of $4\!\times\!4$-matrices
$$
\underline\K_0(t)=\{\underline U\,:\,U\in\K_0(t)\}.
$$
Evidently $\underline\K_0(t_{(k)})$, ($k=1,2,3,4$) are sets of matrices
with all zero elements except elements in the left upper
$k\!\times\!k$-block.

Considering the sets of differential forms
$$
L(t_{(k)})=\{U\in\K_0(t_{(k)})\,:\,U^\dagger=-U\}
$$
and the corresponding sets of matrices $\underline L(t_{(k)})$ as real
Lie algebrae, we see that
$$
L(t_{(k)})\simeq\underline L(t_{(k)})\simeq {\rm u}(k)\simeq
{\rm u}(1)\oplus{\rm su}(k),
$$
where ${\rm u}(k)$, ($k=1,2,3,4$) are Lie algebrae of anti-Hermitian
$k\!\times\!k$-matrices, ${\rm su}(k)$ are the Lie algebrae of traceless
anti-Hermitian matrices, and the sign $\simeq$ denote isomorphism. We
have the Lie groups
$$
G(t_{(k)})\simeq\underline G(t_{(k)})\simeq
{\rm U}(1)\oplus{\rm SU}(k),
$$
where ${\rm SU}(k)$ are the Lie groups of unitary
$k\!\times\!k$-matrices with determinants equal to $1$.

For elements of $L(t_{(k)})$ we define the normalized scalar product
$$
(u,v)_{(k)} := \frac{4}{k} (u,v)
$$
such that
$$
(it_{(k)},it_{(k)})_{(k)}=1,\quad k=1,2,3,4.
$$
Now we show that for every $k=1,2,3,4$ we may
take generators $\tau$ of $L(t_{(k)})$ such that
\begin{equation}
\D_\mu\tau_n=0,\quad(\tau_n,\tau^m)_{(k)}=\delta^m_n,\quad
\tau^\dagger_n=-\tau_n,\quad [\tau_n,\tau_l]=c_{nl}^m\tau_m,
\label{tau}
\end{equation}
where $\tau_n=\tau^n$ and
$c_{nl}^m$ are real structure constants of the Lie algebra
$L(t_{(k)})$.

We use differential forms
\begin{eqnarray*}
\lambda_1 &=& Y^1_2 + Y^2_1\\
\lambda_2 &=& -i Y^1_2 + i Y^2_1\\
\lambda_3 &=& Y^1_1 - Y^2_2\\
\lambda_4 &=& Y^1_3 + Y^3_1\\
\lambda_5 &=& -i Y^1_3 + i Y^3_1\\
\lambda_6 &=& Y^2_3 + Y^3_2\\
\lambda_7 &=& -i Y^2_3 + i Y^3_2\\
\lambda_8 &=& \frac{1}{\sqrt3} (Y^1_1 + Y^2_2 - 2 Y^3_3)
\end{eqnarray*}
such that $\{\underline\lambda_1,\ldots,\underline\lambda_8\}$ is
equivalent to the Gell-Mann basis of the real Lie algebra ${\rm su}(3)$.
We take the following generators of $L(t_{(k)})$:
\medskip

\noindent 1. For $L(t_{(1)}) \simeq {\rm u}(1)$
$$
\tau_0=it_{(1)}.
$$
\noindent 2. For $L(t_{(2)})\simeq {\rm u}(1)\oplus{\rm su}(2)$
$$
\tau_0=it_{(2)},\quad \tau_n=i\lambda_n,\quad
n=1,2,3.
$$
\noindent 3. For $L(t_{(3)})\simeq {\rm u}(1)\oplus{\rm su}(3)$
$$
\tau_0=it_{(3)},\quad \tau_n=i\sqrt{\frac{3}{2}}\lambda_n,\quad
n=1,\ldots,8.
$$
\noindent 4. For $L(t_{(4)})\simeq {\rm u}(1)\oplus{\rm su}(4)$
we take as a basis
$\tau_0,\ldots,\tau_{15}$ the anti-Hermitian basis (\ref{basis:ah})
\medskip

Therefore,
$$
L(t_{(k)})=\{f_n\tau^n\},
$$
where $f_n=f_n(x)$ are real valued scalar functions.

Let us define the set of differential forms
$$
\underbrace{L(t)}=\{U\in L(t)\,:\,\Tr U=0\}.
$$
We see that
$$
\underbrace{L(t_{(k)})}\simeq{\rm su}(k),\quad k=2,3,4.
$$
If we replace $L(t)$ by $\underbrace{L(t)}$ in above considerations,
then we get that all results are valid (we must take
$J^\mu=i\Psi^\dagger\alpha^\mu\Psi-\Tr(i\Psi^\dagger\alpha^\mu\Psi)$ in
Theorem 6 and in eq. (\ref{J})).


\section{Special cases}
Let us denote
$$
I=-ie^{12}.
$$
Then
$$
t_{(1)}=\frac{1}{4}(1+H)(1-iI)
$$
and
$$
it_{(1)}=It_{(1)}.
$$
\theorem 10. For a given $\Phi\in\I(t_{(1)})$ the equation
\begin{equation}
\Psi t_{(1)}=\Phi,
\label{one-to-one}
\end{equation}
has a unique solution $\Psi\in\Lambda_{\even}$.
\par

\proof. We have the orthonormal basis of $\I(t_{(1)})$
$$
t_\k=F_\k t_{(1)},\quad {\textsc k}=1,2,3,4,
$$
where $F_\k\in\Lambda_\even$. Decomposing $\Phi\in\I(t_{(1)})$ with
respect to the basis $t_\k$
\begin{equation}
\Phi=(\alpha^\k+i\beta^\k)t_\k,\quad \alpha^\k,\beta^\k\in\Lambda_0,
\label{Phi}
\end{equation}
and using the relation $it_{(1)}=It_{(1)}$, we see that the differential
form
$$
\Psi=F_\k(\alpha^\k+I\beta^\k)\in\Lambda_\even
$$
is a solution of eq. (\ref{one-to-one}).

We claim that if
$$
U=u+u_{01}e^{01}+u_{02}e^{02}+u_{03}e^{03}+u_{12}e^{12}+u_{13}e^{13}+
u_{23}e^{23}+u_{0123}e^{0123}\in\Lambda_\even
$$
is a solution of the homogeneous equation
$$
Ut_{(1)}=0,
$$
then $U=0$. Indeed, the differential form $Ut_{(1)}\in\I(t_{(1)})$ can
be decomposed into the basis $t_\k$
$$
Ut_{(1)}=\frac{1}{2}(u-iu_{12})t_1+\frac{1}{2}(-u_{13}-iu_{23})t_2+\\
\frac{1}{2}(u_{03}-iu_{0123})t_3+\frac{1}{2}(u_{01}+iu_{02})t_4.
$$
Thus the identity $Ut_{(1)}=0$ implies $U=0$. So the solution
(\ref{Phi}) of eq. (\ref{one-to-one}) is unique. These completes the
proof.
\medskip

\theorem 11. For a given $\Phi\in\I(t_{(2)})$ the equation
\begin{equation}
\Psi t_{(2)}=\Phi,
\label{one-to-one2}
\end{equation}
has a unique solution $\Psi\in\Lambda^\C_{\even}$.
\par

\proof. We have the orthonormal basis of $\I(t_{(2)})$
$$
t_\k=F_\k t_{(2)},\quad {\textsc k}=1,\ldots,8,
$$
where $F_\k\in\Lambda^\C_\even$. Decomposing $\Phi\in\I(t_{(2)})$ with
respect to the basis $t_\k$
\begin{equation}
\Psi=(\alpha^\k+i\beta^\k)t_\k,\quad \alpha^\k,\beta^\k\in\Lambda_0,
\label{Phi2}
\end{equation}
we see that the differential
form
$$
\Psi=F_\k(\alpha^\k+i\beta^\k)\in\Lambda^\C_\even
$$
is a solution of eq. (\ref{one-to-one2}).

We claim that if
$$
U=(v+iw)+\sum_{0\leq
a<b\leq3}(v_{ab}+iw_{ab})e^{ab}+(v_{0123}+iw_{0123})e^{0123}
$$
is a solution of the homogeneous equation
$$
Ut_{(2)}=0,
$$
then $U=0$. Indeed, the differential form $Ut_{(2)}\in\I(t_{(2)})$ can
be decomposed into the basis $t_\k$
$$
Ut_{(2)}=\phi_\k t^\k,\quad \phi_\k=(t_\k,Ut_{(2)}).
$$
We get
\begin{eqnarray*}
\phi_1&=&(v_{}-iv_{12}+iw_{}+w_{12})/2\\
\phi_2&=&(-v_{13}-iv_{23}-iw_{13}+w_{23})/2\\
\phi_3&=&(v_{03}-iv_{0123}+iw_{03}+w_{0123})/2\\
\phi_4&=&(v_{01}+iv_{02}+iw_{01}-w_{02})/2\\
\phi_5&=&(v_{13}-iv_{23}+iw_{13}+w_{23})/2\\
\phi_6&=&(v_{}+iv_{12}+iw_{}-w_{12})/2\\
\phi_7&=&(v_{01}-iv_{02}+iw_{01}+w_{02})/2\\
\phi_8&=&(-v_{03}-iv_{0123}-iw_{03}+w_{0123})/2.
\end{eqnarray*}
Evidently the identity $Ut_{(2)}=0$ implies $U=0$. So the solution
(\ref{Phi2}) of eq. (\ref{one-to-one2}) is unique. These completes the
proof.
\medskip

Denote
$$
\stackrel1 L(t)=\{U\in\Lambda_\even\,:\,U^\dagger=-U, [U,t_{(1)}]=0\}.
$$
The Lie algebra $\stackrel1 L(t)$ is isomorphic to the Lie algebra ${\rm
u}(1)$ and as a generator of $\stackrel1 L(t)$ we may take $\tau_0=I$.

\theorem 12. Differential forms $\stackrel1\Psi\in\Lambda_\even$, $\stackrel1
A_\mu\in\stackrel1 L(t)\top_1$ satisfy the equation
\begin{equation}
dx^\mu(\D_\mu\stackrel1\Psi+\stackrel1\Psi\stackrel1 A_\mu+B_\mu\stackrel1\Psi)H+m\stackrel1\Psi I=0
\label{even:eq}
\end{equation}
iff differential forms $\Psi=\stackrel1\Psi t_{(1)}\in\I(t_{(1)})$,
$A_\mu=\stackrel1 A_\mu t_{(1)}\in L(t)\top_1$ satisfy eq. (\ref{Dirac:type}).
\par

\proof. Multiplying (\ref{even:eq}) from the right by $t_{(1)}$  and
using relations $H t_{(1)}=t_{(1)}$, $I t_{(1)}=it_{(1)}$, we obtain
that $\Psi=\stackrel1\Psi t_{(1)}$,
$A_\mu=\stackrel1 A_\mu t_{(1)}$ satisfy (\ref{Dirac:type}).

Conversely, let $\Psi\in\I(t)$, $A_\mu\in L(t)\top_1$ satisfy
(\ref{Dirac:type}). By Theorem 10 there exists a unique solution
$\stackrel1\Psi\in\Lambda_\even$, $\stackrel1 A_\mu\in\Lambda_\even\top_1$ of the
system of equations $\stackrel1\Psi t_{(1)}=\Psi$, $\stackrel1 A_\mu t_{(1)}=A_\mu$.
It can be shown that $\stackrel1 A_\mu\in\stackrel1 L(t)\top_1$. Substituting
$\stackrel1\Psi t_{(1)}, \stackrel1 A_\mu t_{(1)}$ for $\Psi,A_\mu$ in
(\ref{Dirac:type}), we arrive at the equality
$$
(dx^\mu(\D_\mu\stackrel1\Psi+\stackrel1\Psi\stackrel1 A_\mu+B_\mu\stackrel1\Psi)H+m\stackrel1\Psi
I)t_{(1)}=0,
$$
which we rewrite as
$$
\stackrel1\Omega t_{(1)}=0.
$$
We see that $\stackrel1\Omega\in\Lambda_\even$ and, according to Theorem 10, we
get
$$
\stackrel1\Omega=0.
$$
These completes the proof.
\medskip

Denote
$$
\stackrel2 L(t)=\{U\in\Lambda^\C_\even\,:\,U^\dagger=-U, [U,t_{(2)}]=0\}.
$$
The Lie algebra $\stackrel2 L(t)$ is isomorphic to the Lie algebra ${\rm
u}(1)\oplus{\rm su}(2)$ and we may take the following generators of
$\stackrel2 L(t)$:
$$
\tau_0=i,\quad
\tau_1=e^{23},\quad
\tau_2=-e^{13},\quad
\tau_3=e^{12}.
$$

\theorem 13. Differential forms $\stackrel2\Psi\in\Lambda^\C_\even$, $\stackrel2
A_\mu\in\stackrel2 L(t)\top_1$ satisfy the equation
\begin{equation}
dx^\mu(\D_\mu\stackrel2\Psi+\stackrel2\Psi\stackrel2 A_\mu+B_\mu\stackrel2\Psi)H+im\stackrel2\Psi=0
\label{even:C}
\end{equation}
iff differential forms $\Psi=\stackrel2\Psi t_{(2)}\in\I(t_{(2)})$,
$A_\mu=\stackrel2 A_\mu t_{(2)}\in L(t)\top_1$ satisfy eq. (\ref{Dirac:type}).
\par

\proof\, is word for word identical to the proof of the preceding theorem.
\medskip

Equations (\ref{even:eq}),(\ref{even:C}) together with correspondent
Yang-Mills equations were considered in \cite{nona}-\cite{nona2}.


\end{document}